\documentclass[a4paper,11pt]{article}
\usepackage{graphicx}
\pdfoutput=1 

\usepackage{jinstpub} 

\usepackage[english]{babel}

\usepackage[separate-uncertainty,exponent-product=\cdot]{siunitx}
\DeclareSIUnit{\neq}{n_{eq}}
\DeclareSIUnit{\electron}{e^-}
\DeclareSIUnit{\arbitraryunit}{a.u.}
\DeclareSIUnit{\micron}{\micro\meter}
\DeclareSIUnit{\microns}{\micro\meter}
\DeclareSIUnit{\ppcm}{\si{p\per\centi\meter\squared}}
\DeclareSIUnit{\neqpcm}{\si{n_{eq}\per\centi\meter\squared}}

\usepackage{caption}
\usepackage{subcaption}

\usepackage{floatrow}
\newfloatcommand{capbtabbox}{table}[][\FBwidth]

\usepackage{orcidlink}

\usepackage{tikz}
\usepackage{TikZ/tikzit}

\tikzstyle{blue}=[-, fill={rgb,255: red,121; green,150; blue,209}, draw={rgb,255: red,121; green,150; blue,209}]
\tikzstyle{green}=[-, fill={rgb,255: red,186; green,215; blue,0}, draw={rgb,255: red,186; green,215; blue,0}]
\tikzstyle{brown}=[-, fill={rgb,255: red,199; green,141; blue,53}, draw={rgb,255: red,199; green,141; blue,53}]
\tikzstyle{purple}=[-, fill={rgb,255: red,204; green,81; blue,171}, draw={rgb,255: red,204; green,81; blue,171}]
\tikzstyle{implant}=[-, fill={rgb,255: red,254; green,179; blue,117}, draw={rgb,255: red,254; green,179; blue,117}]
\tikzstyle{epi}=[-, fill={rgb,255: red,255; green,220; blue,204}, draw={rgb,255: red,255; green,220; blue,204}]
\tikzstyle{substrate}=[-, fill={rgb,255: red,252; green,138; blue,100}, draw={rgb,255: red,252; green,138; blue,100}]
\tikzstyle{n_implant}=[-, fill={rgb,255: red,243; green,108; blue,54}, draw={rgb,255: red,243; green,108; blue,54}]
\tikzstyle{backside_metal}=[-, fill={rgb,255: red,199; green,141; blue,53}, draw={rgb,255: red,199; green,141; blue,53}]
\tikzstyle{double_arrow}=[<->, fill=none, line width=0.5mm]
\tikzstyle{single_arrow}=[fill=none, ->, line width=0.5mm]

\usepackage{hyperref} 
\newcommand{\customref}[2]{%
    \hyperref[#2]{\textcolor{blue}{#1~\ref{#2}}}%
}

\title{\boldmath Characterisation of Crystalline Defects in 4H Silicon Carbide using DLTS and TSC}

\author[a,b,1]{Niels G. Sorgenfrei\note{Corresponding author.},\orcidlink{0000-0002-5729-6004}}
\author[c]{Elias Arnqvist,\orcidlink{0009-0001-0029-9757}}
\author[b]{Yana Gurimskaya,\orcidlink{0000-0002-2549-4153}}
\author[b]{Michael Moll,\orcidlink{0000-0001-7013-9751}}
\author[a]{Ulrich Parzefall,}
\author[b]{Faiza Rizwan,}
\author[b]{Moritz Wiehe\orcidlink{0000-0002-6023-8753}}

\affiliation[a]{Institute of Physics, Albert-Ludwigs-Universitaet Freiburg,\\Hermann-Herder-Strasse 3, Freiburg im Breisgau, Germany}
\affiliation[b]{CERN, European Organization for Nuclear Research,\\Esplanade des Particules 1, Geneva, 1211, Switzerland}
\affiliation[c]{Department of Physics and Astronomy,\\Uppsala University, Box 516, Uppsala, 75137, Sweden}

\emailAdd{niels.sorgenfrei@cern.ch}

\abstract{Future hadron collider experiments will require sensing materials that withstand stronger radiation fields. 
Therefore, either a frequent replacement of detectors, a significant increase in radiation hardness of Silicon, or a shift to different materials is needed. 
Wide-bandgap materials are a natural choice, due to their significantly reduced leakage currents, even after irradiation. 
In recent years, substantial progress in the production of high-quality monocrystalline Silicon Carbide of the 4H polytype has led to a renewed interest in this material.

In this article, a study of electrically active defects in a n-type epitaxial 4H Silicon Carbide diode is presented. 
By employing spectroscopical measurement methods, like Deep-Level Transient Spectroscopy (DLTS) and Thermally Stimulated Currents (TSC), energy levels in the bandgap are investigated. 
Defect parameters like concentration, activation energy and capture cross-section are stated. 
A simulation framework was utilised to compare and match the results from the two methods.

This study is made in the context of a study of radiation hardness of 4H Silicon Carbide sensors.
Other studies investigating macroscopic properties of the material, like their charge collection efficiency after irradiation, were performed on the same kind of diodes.

This study provides a first set of measured defect parameters in state-of-the-art 4H-SiC material, from defects present prior to irradiation. 
These defects are intrinsic, such as vacancies, related to impurities and doping imperfections, or are growth related. 
The $Z_\text{1/2}$ defect and a Nitrogen related defect were identified.
}


\arxivnumber{arXiv:2505.02671} 

\proceeding{20$^{\text{th}}$ Anniversary "Trento" Workshop on Advanced Silicon Radiation Detectors\\
  4-6 Feb 2025\\
  FBK, Trento, Italy}

\usepackage{microtype}
\begin{document}
\maketitle 
\flushbottom

\section{Introduction}
Silicon (Si) as a material has been dominating the field of particle tracking detectors in the last decades.
This decades long experience with the material and the performed research on its properties make it the obvious choice for building tracking detectors.
However, Si detectors suffer greatly from radiation damage, and future hadron collider particle detector experiments will require their detectors to withstand even higher radiation fluences.
A lot of effort is being put into increasing the radiation hardness of Si, by means of defect engineering or new detector geometries \cite{MollDisplacement}.
Another path to take is exploring different materials.
Silicon Carbide (SiC) as a sensing material for high energy particle physics (HEP) experiments has gained newfound interest by researchers in the last years.
This interest can be partly attributed to improvements in the SiC material quality.
This is fuelled by industry interest, due to the electrical properties of SiC, favourable for high efficiency power devices.
For HEP experiments, some of the material properties of SiC give indications of a possible increased radiation hardness and ease of operation in a detector system, compared to Si.
As a consequence of the large bandgap of SiC, extremely low leakage currents at elevated temperatures, under ambient light conditions and after irradiation are observed \cite{Rafí}.

In this article, a 4H-SiC n-type diode from IMB-CNM \cite{CNM} is investigated.
The same kind of diodes have been previously studied by other authors to investigate their macroscopic properties like charge collection, current and capacitance characteristics, before and after irradiation \cite{Rafí,Rafi,Gaggl}.
This article adds information about the presence of defects in the material.
The measured defect parameters can be used to simulate the detector performance for these diodes.
A follow up article will report on defect parameters after different types of radiation.

Defect spectroscopy methods, specifically the Thermally Stimulated Current and the Deep-Level Transient Spectroscopy techniques, are used to study the presence of defects.
In-depth explanations about the used measurement methods can be found in Ref. \cite[Chapters 4.4 \& 4.5]{Moll} and the references therein.
The used cryostat can only reach a maximum temperature of \SI{350}{\kelvin}.
To scan the full bandgap of 4H-SiC, temperatures of up to \SI{800}{\kelvin} are required.
Therefore, only a fraction of the full bandgap is shown, and potential defects situated at higher temperatures are missed.

\section{Devices\label{sec:devices}}
\begin{figure}[htbp]
\centering
\resizebox{.5\linewidth}{!}{%
\input{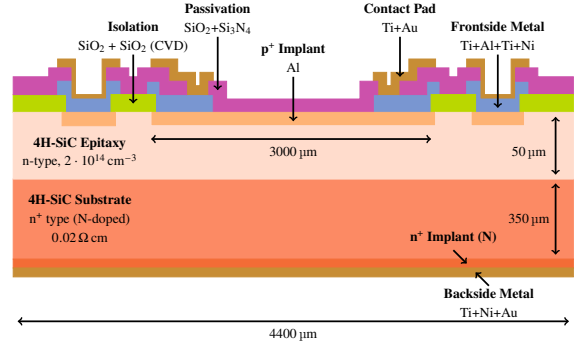}
}
\caption{Schematic, cross-sectional view of the investigated 4H-SiC diode, adapted from the producer's \cite{CNM} schematic. The feature-size is not to scale.}
    \label{fig:50micron-schematic}
\end{figure}

The diode under investigation is produced by IMB-CNM \cite{CNM}.
The wafer was bought from Ascatron AB (later acquired by II-VI/Coherent Corp.).
A schematic drawing of the diode is shown in \autoref{fig:50micron-schematic}.
Articles on similar structures can be found in Ref. \cite{Rafí,Rafi,Gaggl}.
The diode has an active, n-doped, epitaxially grown layer of \SI{50}{\micron} 4H-SiC on top of a \SI{350}{\microns} thick, highly doped substrate wafer.
The n-type doping is achieved using Nitrogen.
The active area is \SI{0.1}{\centi\meter\squared}.
From Capacitance-Voltage measurements, the effective doping concentration of the epi layer is calculated to be \SI{1.6e14}{\per\cubic\centi\meter}.

\section{Thermally Stimulated Currents (TSC)}
\begin{figure}[htbp]
    \centering
    \begin{subfigure}[t]{.49\linewidth}
        \centering
        \includegraphics[width=.9\linewidth]{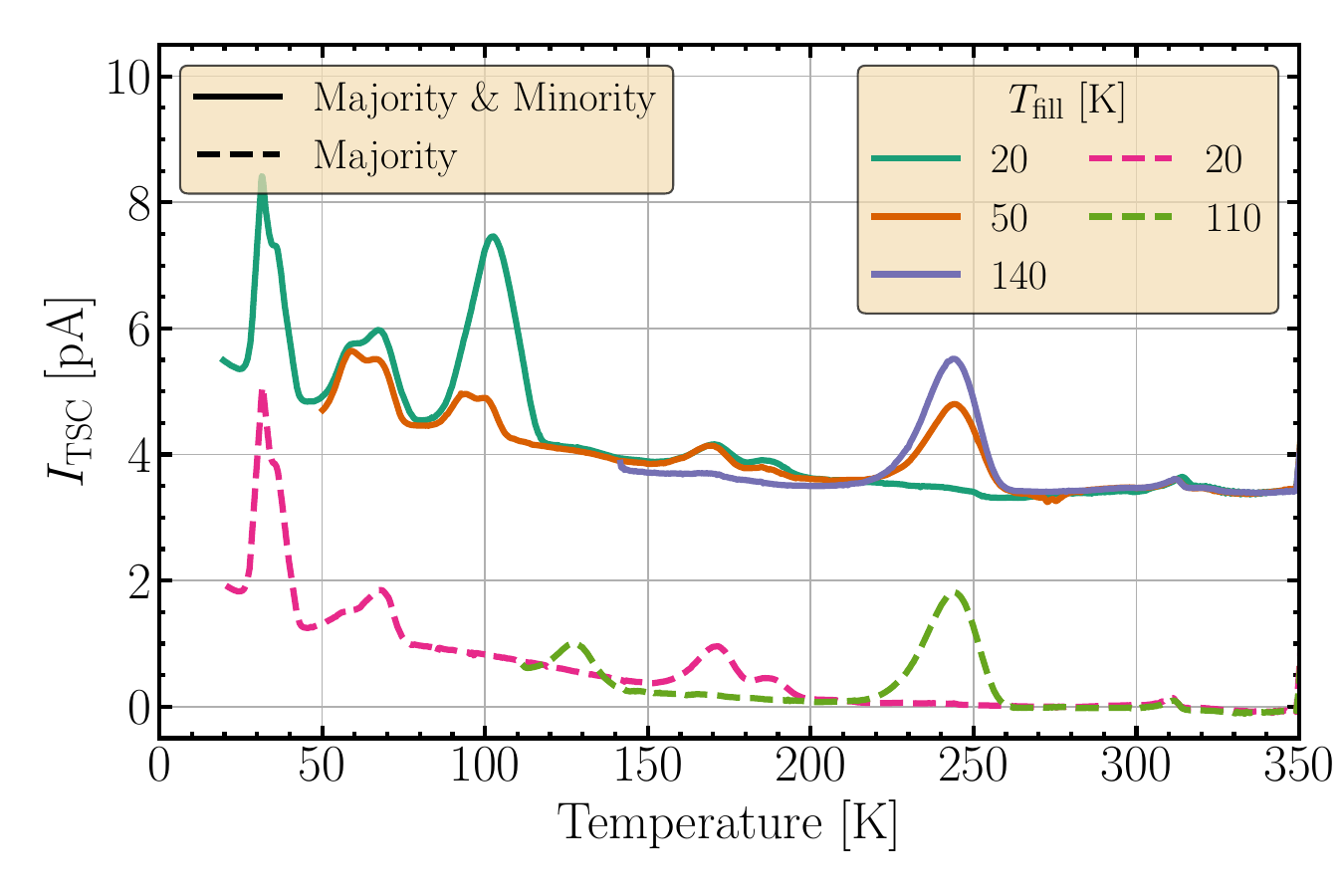}
        \caption{TSC Spectra}
        \label{fig:tsc-temp-combined}
    \end{subfigure}
    \hfill
    \begin{subfigure}[t]{.49\linewidth}
        \centering
       \includegraphics[width=.9\linewidth]{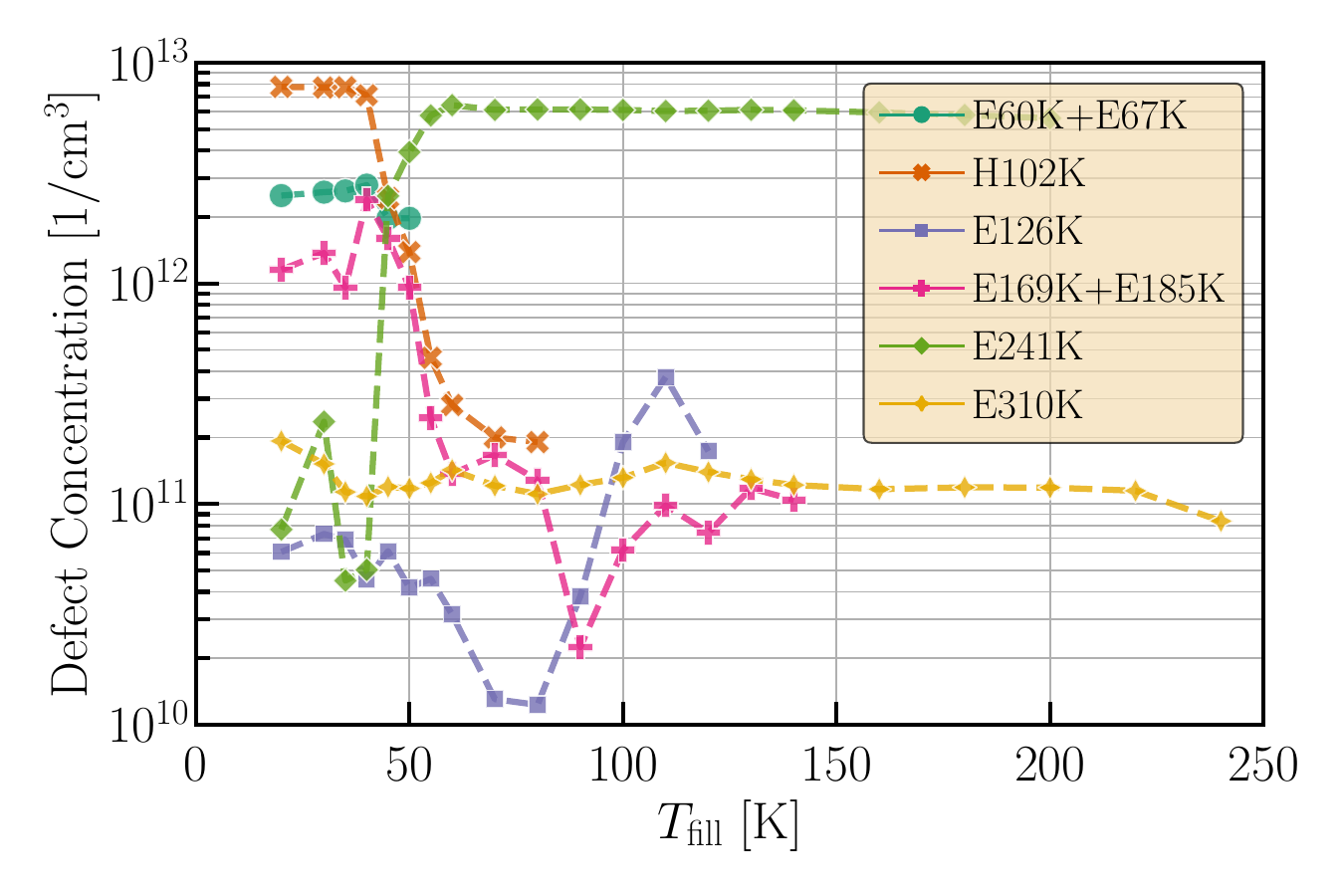}
    \caption{Defect Concentrations}
    \label{fig:tsc-defect_concentrations}
    \end{subfigure}
    \caption{(\subref{fig:tsc-temp-combined}) TSC spectra for different $T_\text{fill}$. Solid lines indicate forward biased and dashed lines \SI{0}{\volt} filling conditions. The spectra have been shifted for the sake of clarity. (\subref{fig:tsc-defect_concentrations}) Defect concentrations versus $T_\text{fill}$, extracted from the forward biased filling measurements (majority \& minority carriers).}
\end{figure}

The results from TSC measurements are displayed in \autoref{fig:tsc-temp-combined}, for different filling temperatures $T_\mathrm{fill}$.
11 defects are observed in the temperature range of \SIrange{20}{350}{\kelvin}.
Since the measured diode is unirradiated, all observed defects must be either intrinsic (interstitials, vacancies, anti-sites), extrinsic (impurity and doping (Nitrogen, Aluminium, etc.)) or growth related.
A match of the measured defects with known defects from literature is possible for the $Z_\text{1/2}$ defect, labelled here as \textit{E241K}.
This electron trap is often referred to as the main lifetime killer in SiC \cite{Capan}.

\paragraph{Filling Temperature}
To observe the dependence on $T_\mathrm{fill}$, the diode was repeatedly measured under identical measurement conditions, while only varying $T_\mathrm{fill}$ between \SIlist{20;250}{\kelvin}.
This was measured for forward bias (majority and minority carrier) and \SI{0}{\volt} (majority carrier) filling conditions, see \autoref{fig:tsc-temp-combined} for selected $T_\mathrm{fill}$.
All defects display a dependence on the filling temperature.
Some defects are only filled at certain conditions.
This is also shown in \autoref{fig:tsc-defect_concentrations} where the extracted defect concentrations are plotted versus $T_\mathrm{fill}$.
While some defects located at high temperatures are not filled at low temperatures, some are exclusively filled at low temperatures, e.g. the two defects located between \SIlist{150;200}{\kelvin}. 
While the majority carrier injection shows a good filling at \SIlist{20;30}{\kelvin}, the forward bias filling measurements show an optimal filling temperature of \SI{40}{\kelvin}, with suppressed filling at higher and lower temperatures.
Observing the hole trap \textit{H102K}, located at $\sim\SI{100}{\kelvin}$, reveals a double peak structure, not visible for $T_\mathrm{fill}<\SI{45}{\kelvin}$. 
There are two overlapping trap levels, and one must have a stronger capture cross-section or a larger concentration than the other, completely overshadowing its signal at certain filling conditions.

The $Z_\text{1/2}$ defect exhibits no dependence on $T_\mathrm{fill}$ between \SIlist{60;220}{\kelvin} for the \SI{0}{\volt} filling.
However, it is not being filled at \SI{20}{\kelvin} or \SI{30}{\kelvin}.
An explanation through the amount of available charges (i.e. effective doping concentration) being the limiting factor here is revoked through the observation of the forward biased injection.
There, no limitation in available charge carriers is present.
No filling of this defect at low temperatures is observed as well.
Therefore, it is more plausible that this defect has a very low capture cross-section at low temperatures.

\begin{figure}[htbp]
    \centering
    \begin{subfigure}[t]{.49\linewidth}
        \centering
        \includegraphics[width=.9\linewidth]{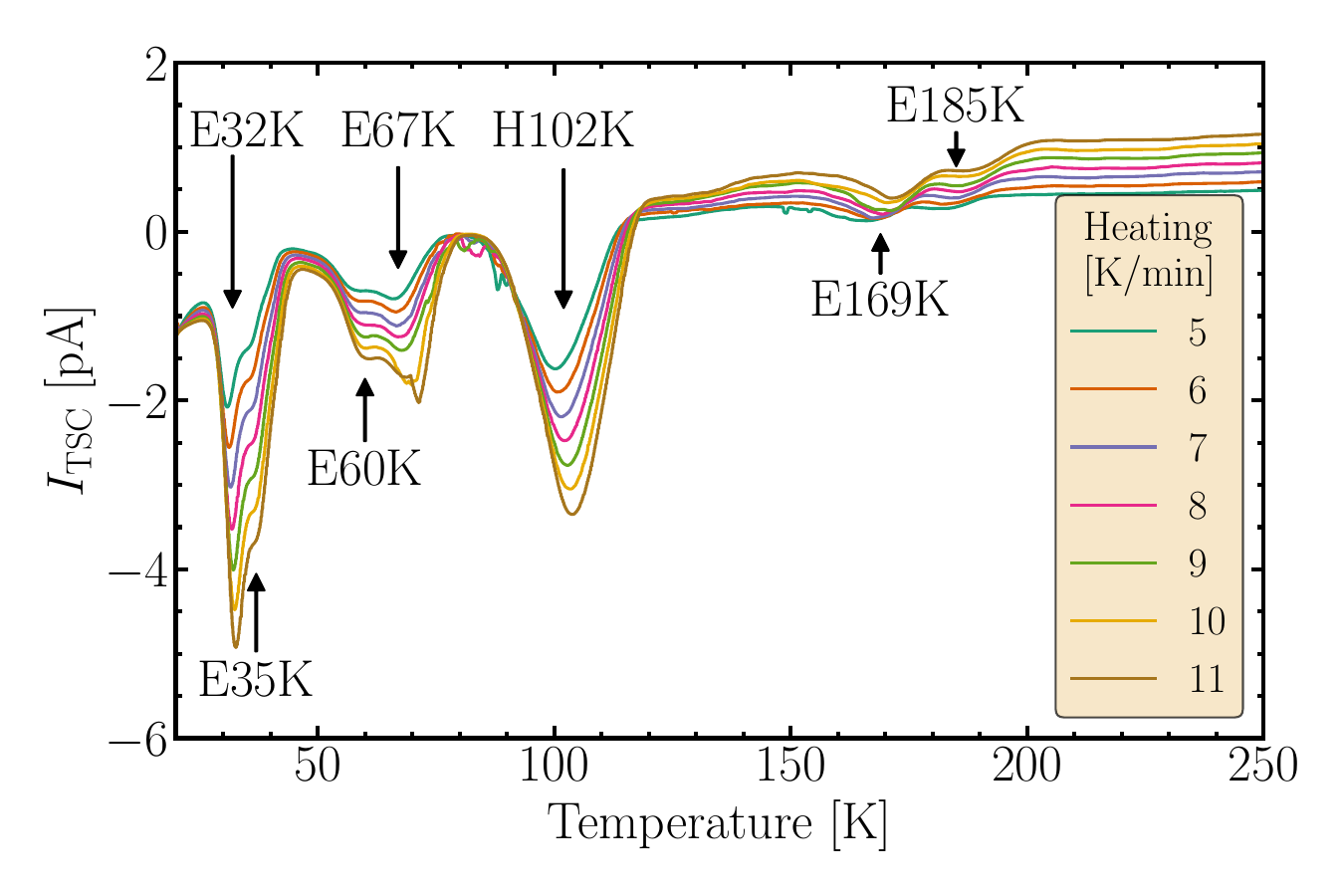}
        \caption{$T_\mathrm{fill}=\SI{20}{\kelvin}$}
        \label{fig:tsc-unirrad-50-HR}
    \end{subfigure}
    \hfill
    \begin{subfigure}[t]{.49\linewidth}
        \centering
        \includegraphics[width=.9\linewidth]{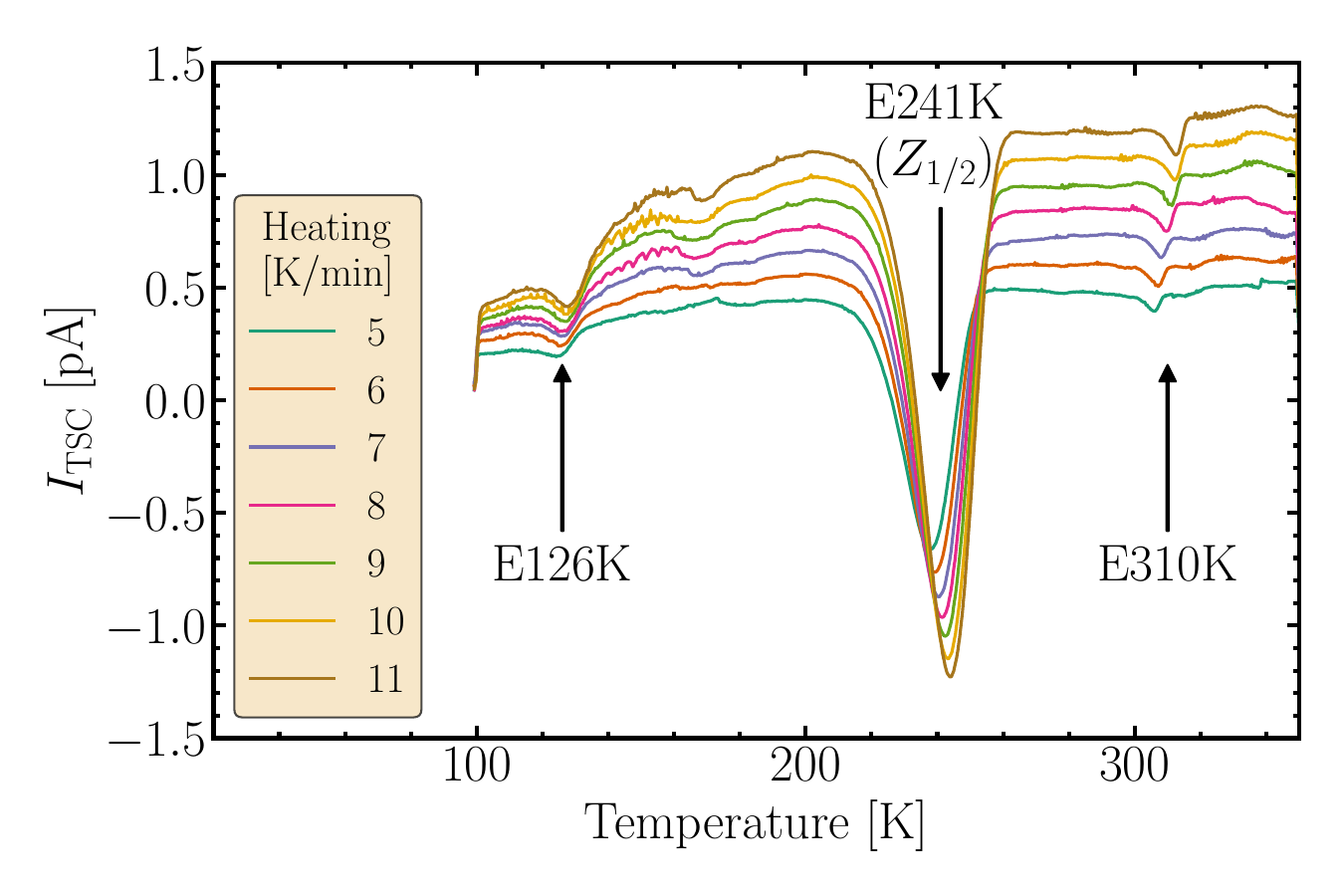}
        \caption{$T_\mathrm{fill}=\SI{100}{\kelvin}$}
        \label{fig:tsc-unirrad-50-HR-100Kfill}
    \end{subfigure}
    \caption{Heating rate scans for (\subref{fig:tsc-unirrad-50-HR}) $T_\mathrm{fill}=\SI{20}{\kelvin}$ and (\subref{fig:tsc-unirrad-50-HR-100Kfill}) \SI{100}{\kelvin}.}
    \label{fig:TSC_meas}
\end{figure}

\begin{figure}[htbp]
\begin{floatrow}
\ffigbox{%
    \centering
    \includegraphics[width=\linewidth]{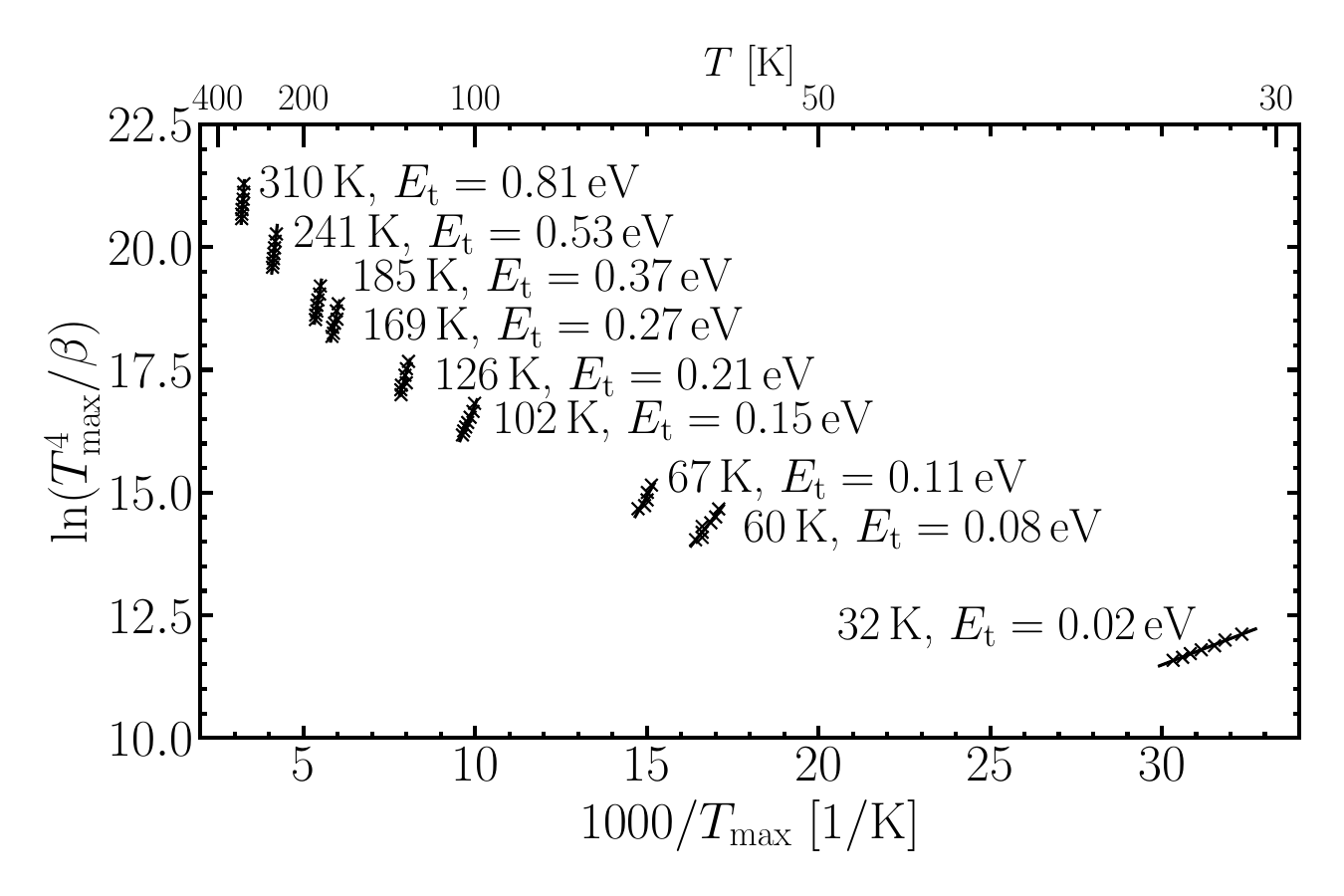}
}{%
   \caption{Arrhenius plot from the measurements in \autoref{fig:tsc-unirrad-50-HR} and \subref{fig:tsc-unirrad-50-HR-100Kfill}. The black lines represent linear fits onto the data points, used to obtain the energy from their slopes.}
    \label{fig:tsc-unirrad-50-Arr}
}
\hfill
\capbtabbox{%
\centering
\resizebox{.95\linewidth}{!}{%
\begin{tabular}{lccccc}
Defect & Concentration                 & Energy & e & h & Bias dep-    \\
name   & [\si{\per\centi\meter\cubed}] & [eV]   &   &   & endant shift \\\hline
\textit{E32K}  & -            & 0.02 & x &   & no  \\
\textit{E35K}  & -            & -    & x &   & no  \\
\textit{E60K}  & -            & 0.08 & x &   & yes \\
\textit{E67K}  & -            & 0.11 & x &   & no  \\
\textit{H102K} & \num{7.8e12} & 0.15 &   & x & yes \\
\textit{E126K} & \num{6.1e11} & 0.21 & x &   & no  \\
\textit{E169K} & \num{1.6e12} & 0.27 & x &   & no  \\
\textit{E185K} & \num{1.1e12} & 0.37 & x &   & no  \\
\textit{E241K} & \num{6.4e12} & 0.53 & x &   & no  \\
\textit{E310K} & \num{1.9e11} & 0.81 & x &   & no  \\
\end{tabular}
}}{%
  \caption{Defect parameters from TSC measurements. The e \& h columns indicate electron or hole traps. The last column indicates if the peaks' positions depend on the reverse bias.}
\label{tab:parameters-50-unirrad}
}
\end{floatrow}
\end{figure}

\paragraph{Heating Rate Measurements}
By measuring at various heating rates, the energy levels of the defects can be extracted. 
The results are shown in \autoref{fig:TSC_meas}. An Arrhenius relation is used to calculate the energies from the shift of the peaks’ positions (see \autoref{fig:tsc-unirrad-50-Arr}), following the procedure described in Ref. \cite{Fang}. 
The energy values always indicate the relative energetic distance to either the valence or conduction band.
It is not possible to extract the energy for \textit{E35K}, since it is overlapping to such an extent with \textit{E32K}, that a discernible extrema is not readily apparent.
The (low) accuracy of the heating rate method is discussed in \autoref{sec:comparison}.

\customref{Table}{tab:parameters-50-unirrad} lists the complete set of extracted defect parameters from the previously described measurements.
For overlapping defects, an extraction of their concentration is not possible.

\begin{figure}[htbp]
\centering
\begin{subfigure}[t]{0.56\linewidth}
    \centering
    \includegraphics[width=\linewidth]{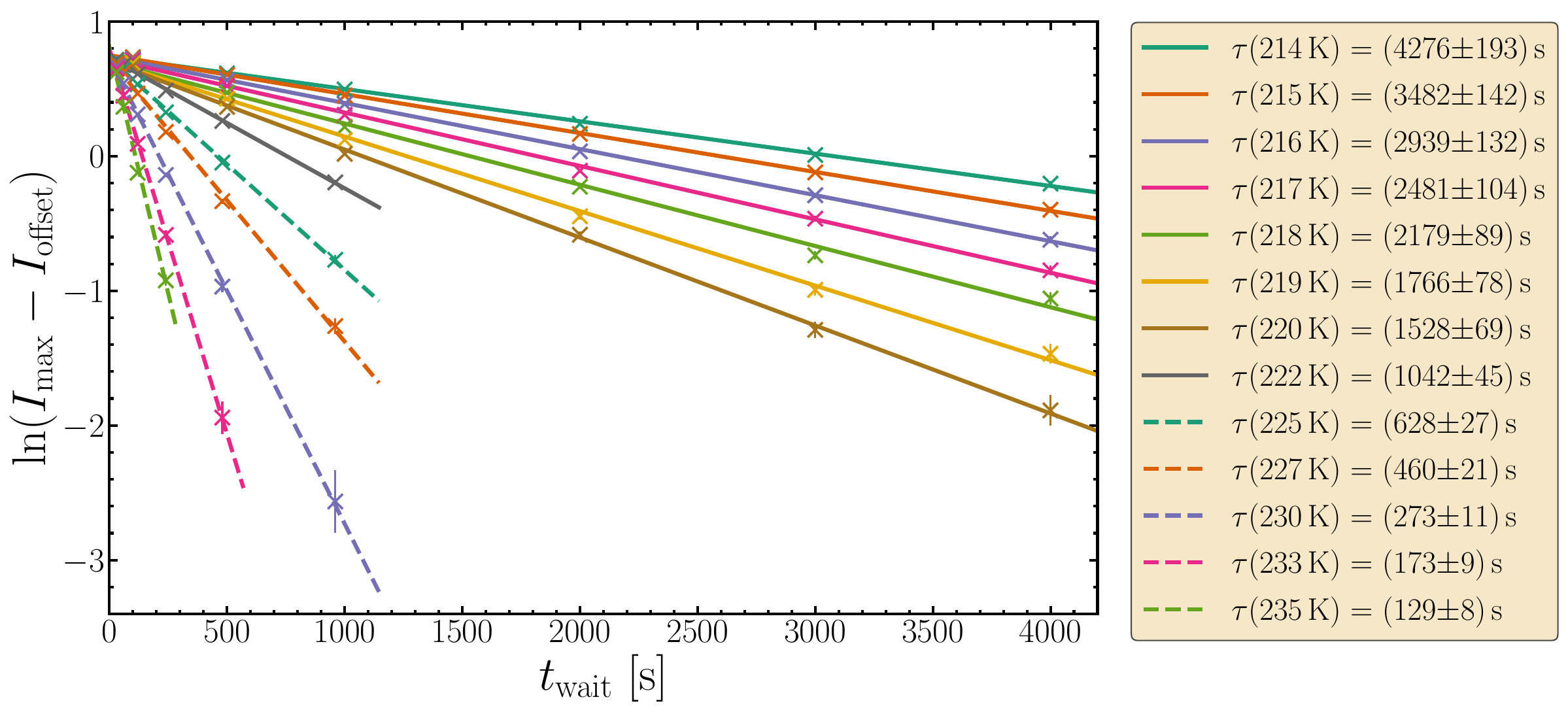}
    \caption{Time Constants}
    \label{fig:time-const_Z12}
\end{subfigure}
\hfill
\begin{subfigure}[t]{0.42\linewidth}
    \centering
    \includegraphics[width=.85\linewidth]{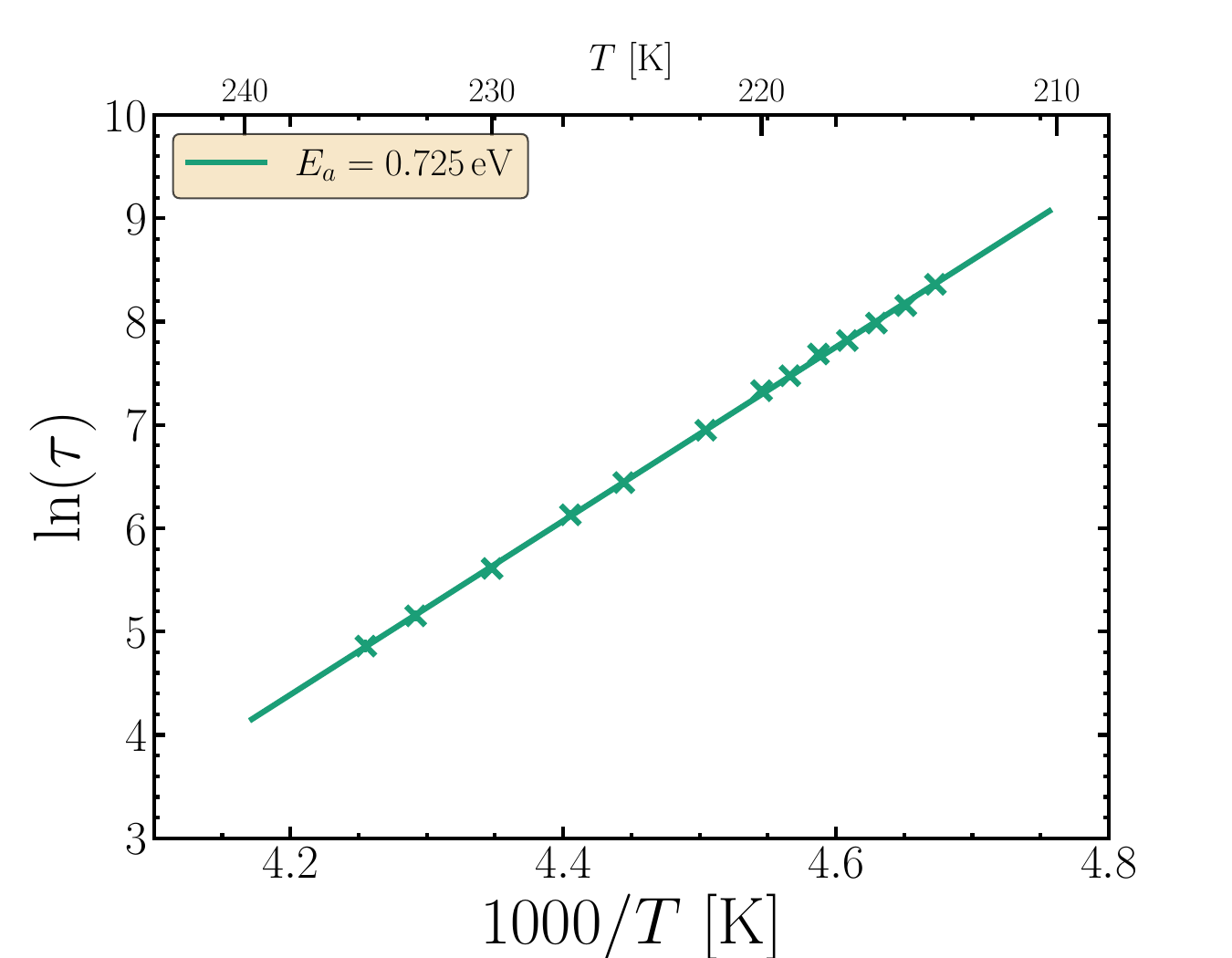}
    \caption{Arrhenius Plot}
    \label{fig:Arrhenius_Z12}
\end{subfigure}
\caption{(\subref{fig:time-const_Z12}) Extraction of the time constants $\tau$ for different filling temperatures. The maximum current released from the $Z_\text{1/2}$ defect is displayed as a function of $t_\mathrm{wait}$. (\subref{fig:Arrhenius_Z12}) Logarithmic display of the extracted time constants as a function of the reciprocal filling temperature.}
\label{fig:delayed_heating}
\end{figure}

\paragraph{Delayed Heating Measurements}
For a more precise determination of the energy level, a delayed heating measurement was performed for the $Z_\text{1/2}$ defect (\textit{E241K}).
The waiting time $t_\text{wait}$, spanning from the end of charge injection to heating initiation, is varied between \SIrange{10}{4000}{\second}. 
This allows for an increased emission of charge carriers, resulting in a reduction of the peak current $I_\text{max}$ in the TSC spectrum.
$T_\text{fill}$ is chosen to be slightly below the temperature of the peak's maximum of the investigated defect.
For each chosen $T_\text{fill}$, TSC spectra are recorded for every $t_\text{wait}$.
From each measurement, $I_\text{max}$ is extracted and the logarithm of its value, corrected for a common offset value $I_\text{offset}$, is plotted against the waiting time to obtain a linear relation (\autoref{fig:time-const_Z12}).
The inverse slope of a linear regression is used to obtain the time constant $\tau$ of the de-trapping process at each filling temperature.
This set of $\tau$ values follows an Arrhenius relation and in a linearised form yields the energy level of this defect from the slope of a linear regression (\autoref{fig:Arrhenius_Z12}).
An uncertainty of \SI{5}{\percent} is estimated for the energy value.
This combines fit uncertainties from both the time constants and the Arrhenius relation, the uncertainty on the extraction of the peak currents and includes estimates on the stability of the temperature ramp.

The results from the measurements of the $Z_\text{1/2}$ defect are shown in 
\autoref{fig:delayed_heating}.
For both plots the expected linear behaviour is observed.
The extracted value for the energy level of the $Z_\text{1/2}$ defect is $\left(\num{0.725(36)}\right)\,\si{\electronvolt}$. 
This value has an uncertainty of roughly \SI{5}{\percent} and deviates by about $\SI{1.8}{\sigma}$ from the one obtained from DLTS measurements, see \customref{Table}{tab:dlts_50_unirrad_parameters}, and $\SI{1.5}{\sigma}$ from the usually quoted value of \SI{0.67}{\electronvolt} in literature, see Ref. \cite{Capan} and references therein.

\section{Deep-Level Transient Spectroscopy (DLTS)}
\begin{figure}[htbp]
    \centering
    \begin{subfigure}[t]{.49\linewidth}
        \centering
        \includegraphics[width=.8\linewidth]{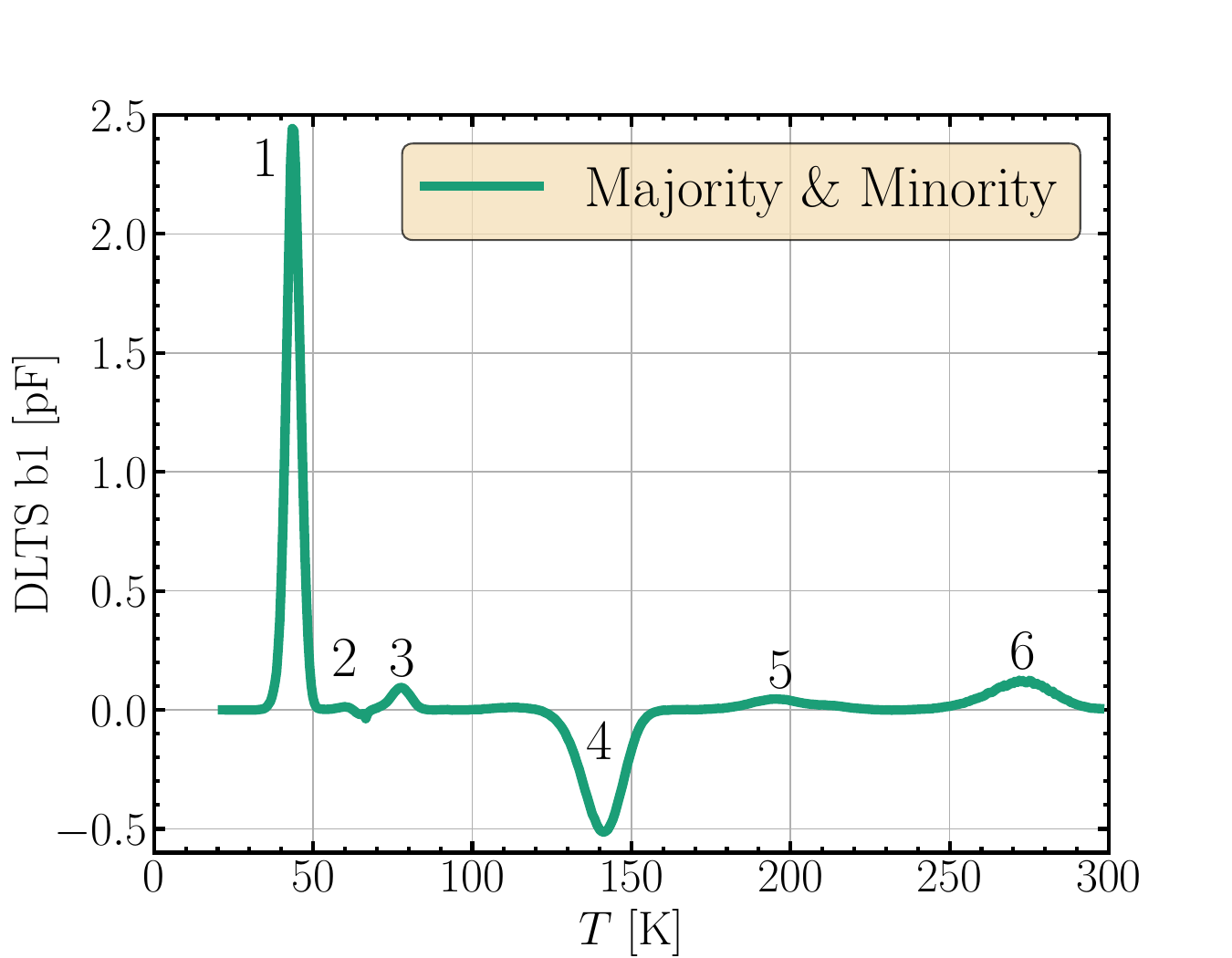}
        \caption{DLTS b1 Spectrum}
        \label{fig:DLTSb1}
    \end{subfigure}
    \hfill
    \begin{subfigure}[t]{.49\linewidth}
        \centering
        \includegraphics[width=.8\linewidth]{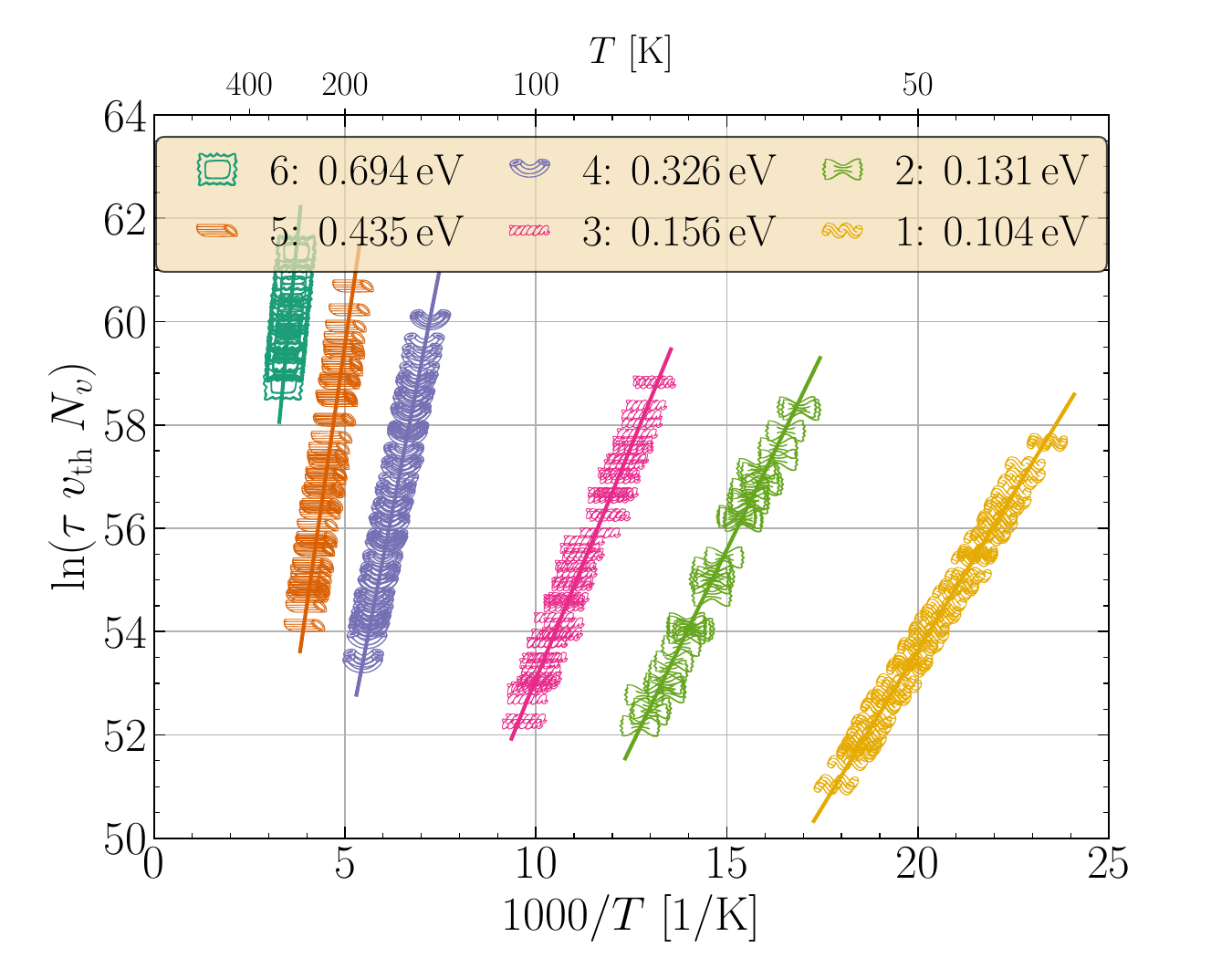}
        \caption{Arrhenius Plot}
        \label{fig:DLTSArrhenius}   
    \end{subfigure}
    \caption{(\subref{fig:DLTSb1}) DLTS spectrum from the b1 correlator function for a rate window of \SI{2}{\second}, a reverse bias of \SI{-10}{\volt} and a forward bias injection with \SI{+3}{\volt} for a pulse duration of \SI{1}{\milli\second}. (\subref{fig:DLTSArrhenius}) Arrhenius plot obtained from the DLTS measurement shown in \autoref{fig:DLTSb1}.}
\end{figure}

\begin{figure}[htbp]
\begin{floatrow}
\ffigbox{%
    \centering
    \includegraphics[width=.75\linewidth]{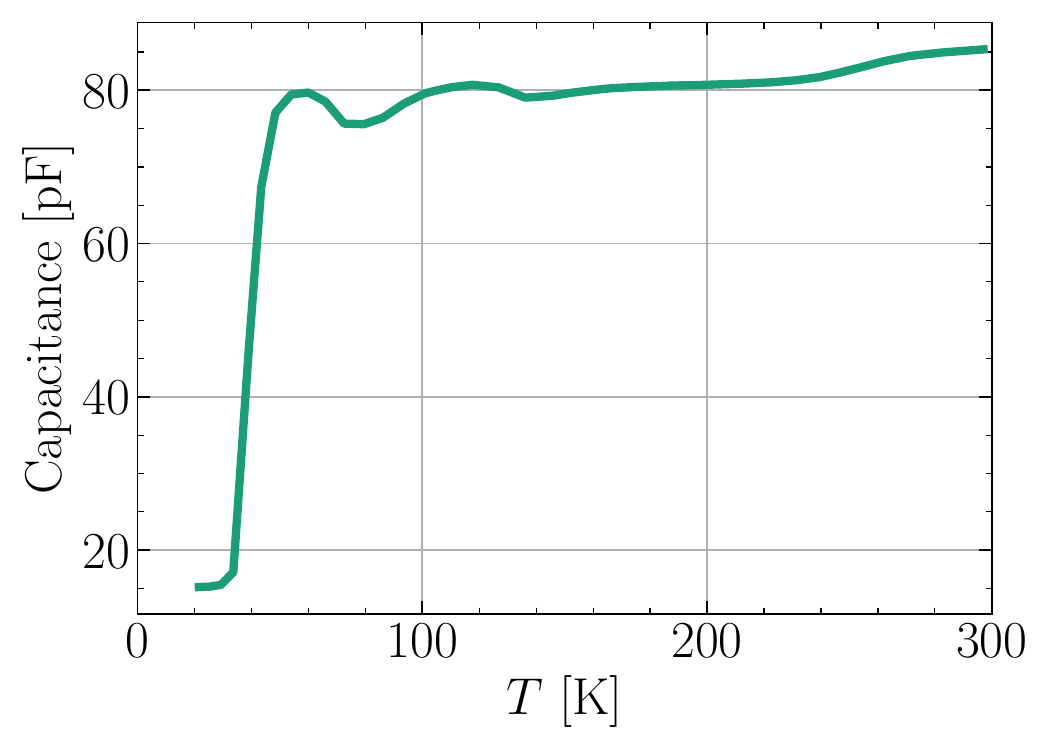}
}{%
   \caption{Reverse capacitance vs. temperature.}
    \label{fig:crev}
}
\hfill
\capbtabbox{%
\centering
\resizebox{.95\linewidth}{!}{%
\begin{tabular}{cccccc}
        Defect & Concentration & Energy & Sigma & e & h \\
        name   & [\si{\per\centi\meter\cubed}] & [\si{\electronvolt}] & [\si{\centi\meter\squared}] & & \\\hline
        1 & \num{7.85e13} & \num{0.104} & \num{1.69e-13} & x &   \\
        2 & \num{3.92e11} & \num{0.131} & \num{5.64e-15} & x &   \\
        3 & \num{1.79e12} & \num{0.156} & \num{6.12e-16} & x &   \\
        4 & \num{1.46e13} & \num{0.326} & \num{6.04e-15} &   & x \\
        5 & \num{2.08e12} & \num{0.435} & \num{1.26e-15} & x &   \\
        6 & \num{3.30e12} & \num{0.694} & \num{2.17e-14} & x &
    \end{tabular}
}
}{%
  \caption{Defect parameters from \autoref{fig:DLTSb1}.}
    \label{tab:dlts_50_unirrad_parameters}
}
\end{floatrow}
\end{figure}

DLTS measurements were carried out with a reverse bias of \SI{-10}{\volt}, a pulse voltage of \SI{-0.6}{\volt} for majority carrier injection and \SI{3}{\volt} for a forward biased majority and minority carrier injection.
The pulses have a duration of \SI{1}{\milli\second} and the capacitance transients are recorded for 3 different rate windows: \SI{20}{\milli\second}, \SI{200}{\milli\second} and \SI{2}{\second}.
The transients are analysed with \num{23} correlator functions see Ref. \cite[Chapter 4.4.3]{Moll}.
The resulting spectrum for a rate window of \SI{2}{\second} and the b1 correlator is displayed in \autoref{fig:DLTSb1}.
An Arrhenius plot for these measurements is displayed in \autoref{fig:DLTSArrhenius}, and the extracted properties of the defects are shown in \customref{Table}{tab:dlts_50_unirrad_parameters}.
The reverse capacitance is displayed in \autoref{fig:crev}.
The capacitance remains stable enough for an accurate extraction of defect parameters above \SI{50}{\kelvin}.
Below that, a drop in capacitance is observed, hindering extraction of precise values.
This is known as dopant freeze-out, which describes the apparent deactivation of dopant atoms due to temperature.
Since the dopant was Nitrogen, defect 1, observed in \autoref{fig:DLTSb1} below \SI{50}{\kelvin}, is assumed to be related to the Nitrogen doping. 
Its concentration of \SI{8e13}{\per\cubic\centi\meter} is nearly half of the manufacture specified doping concentration of \SI{2e14}{\per\cubic\centi\meter}. 
However, the strong drop in capacitance in that temperature region makes an accurate extraction through capacitance-based DLTS impossible, which could explain the difference.
The extracted energy value also fits with in literature quoted values for a Nitrogen substitute on a Carbon cubic lattice site (k-site), see Ref. \cite{nitrogen} and sources therein.

To estimate an uncertainty on the extracted values, the measurements were repeated and user defined parameters in the analysis process were altered to observe their influence. 
The variation of these values for the extracted parameters are of different orders.
The energy value is the most stable one, where the variations stay below \SI{4}{\percent}.
The defect concentration values vary in the order of up to \SI{5}{\percent}, and the capture cross-sections exhibits strong variations of up to \SI{25}{\percent}.

The applied reverse bias of \SI{-10}{\volt} results in a depleted depth of about \SI{9.4}{\microns}.
This means that only defects up to this depth are contributing to the measurements.
The obtained values for the defect concentrations are normalised to the depleted volume.
Assuming a homogeneous defect distribution throughout the bulk, no new defect types should appear or a change in (normalised) concentrations be observed with increased reverse biases.
Repeating the measurements with reverse biases up to \SI{-100}{\volt} ($\approx\SI{26.2}{\micro\meter}$ depleted depth) only showed an increase in the DLTS signal amplitude (due to the larger depleted volume contributing to the signal generation), but no new defect types appeared.

\section{Disscusion of DLTS \& TSC Results\label{sec:comparison}}
Comparing the number of defects and their extracted parameters measured by TSC and DLTS (see \customref{Tables}{tab:parameters-50-unirrad} and \ref{tab:dlts_50_unirrad_parameters}) yields significant differences.
The sometimes large differences in the extracted energy level values between the two measurement methods can be explained by the imprecision inherent in the TSC heating rate method.
To obtain an accurate linear regression from the shift of the defect peak maxima, the range of heating rates must be significantly larger.
Currently, the heating rate is only varied by about a factor of two.
To obtain accurate results, this variation should span three or more orders of magnitude.
However, if the heating rate is increased beyond the normally used \SI{11}{\kelvin\per\minute}, a significant temperature gradient will form within the diode's bulk.
This will lead to distorted TSC spectra due to the inhomogeneous thermally stimulated release of trapped charges at different depths within the sample.
Going to heating rates below \SI{5}{\kelvin\per\minute} results in spectra where the peaks are extremely flat and start to fall within the level of electronic noise.
This limited variation in the heating rate causes the linear regressions to be heavily dependent on statistical fluctuations of the data points.\\\indent
To facilitate comparison and untangle these differences, the custom python framework \textit{pyTSC} \cite{pyTSC} is used to simulate TSC spectra from a set of input defect parameters.
The values for energy levels, defect concentrations and capture cross-sections are taken from DLTS measurements.
The results from these simulations (not shown here) show an agreement between the two measurement methods.
Peak positions and widths agree for the measured defects, only the amplitudes deviate.
These results support, that TSC and DLTS did indeed measure the same defects, albeit TSC a few additional ones.
The TSC heating rate method is simply not precise enough, due to the aforementioned arguments.
In essence, both methods determined the same emission time constants $\tau$, but differ only in their calculation accuracy of the energy levels and capture-cross sections.
The DLTS values for theses parameters are orders of magnitude more precise and reliable, and should be used as the final results from these measurements.

\section{Conclusion\label{sec:conclusion}}
The TSC and DLTS methods were used to perform defect spectroscopy on an unirradiated, n-type, 4H-SiC diode, and revealed a multitude of defects already present in the unirradiated diode.
Therefore, all defects observed must be either intrinsic, doping or impurity related.
Simulations were performed to match the results from DLTS with TSC, which revealed an overall agreement between the two methods, but also that some defects were only observed in either one method.
The defect parameters obtained from the DLTS measurements are of much higher precision and reliability than those from TSC, hence should be perceived as the final results.
Properties of one defect was matched with known literature values, the $Z_\text{1/2}$ defect.
A strong temperature dependence for it's capture cross-section was observed for low temperatures.
One defect was matched to be substitutional Nitrogen on a Carbon cubic lattice-site (k-site).
To further investigate the present defects and ascertain their chemical structure, multiple irradiation campaigns are being carried out to compare damage created by protons, neutrons and gammas.
This might give further insight in the defects observed in this work, in particular giving an understanding if those defects can or cannot be generated by radiation damage and if they are point-like or clustered.
To understand if 4H-SiC detectors are more radiation hard than Si, further studies have to be performed comparing their performance and degradation after radiation.


\appendix

\acknowledgments

This work has been conducted in the framework of the RD50 and DRD3 collaborations and the corresponding author has been sponsored by the Wolfgang Gentner Programme of the German Federal Ministry of Education and Research (grant no. 13E18CHA).

\bibliography{literature}
\bibliographystyle{unsrt}

\end{document}